\documentclass[aps,prl,reprint,nofootinbib,superscriptaddress,preprintnumbers,amsmath,amssymb,latexsym,array,enumerate,letterpaper,twocolumn]{revtex4-1}

\usepackage{inputenc,amsmath,amssymb}
\usepackage{graphicx}
\usepackage{natbib}
\usepackage{xcolor}
\usepackage[colorlinks=true,urlcolor=brown,linkcolor=blue,citecolor=magenta]{hyperref}
\usepackage{graphicx}
\usepackage{comment}

\newcommand{\beq}{\begin{eqnarray}}
\newcommand{\eeq}{\end{eqnarray}}
\newcommand{\bea}{\begin{eqnarray}}
\newcommand{\eea}{\end{eqnarray}}
\newcommand{\bwt}{\begin{widetext}}
\newcommand{\ewt}{\end{widetext}}
\newcommand{\pL}{\left(}
\newcommand{\pR}{\right)}
\newcommand{\eV}{\textrm{eV}}
\newcommand{\meV}{\textrm{meV}}
\newcommand{\MeV}{\textrm{MeV}}

\begin{document}

\title{Dark Solar Wind}

\author{Jae Hyeok Chang}
\email{jaechang@umd.edu}
\affiliation{Department of Physics and Astronomy, Johns Hopkins University, Baltimore, MD 21218, USA}
\affiliation{Maryland Center for Fundamental Physics, Department of Physics, University of Maryland, College Park, MD 20742, USA}

\author{David E. Kaplan}
\email{david.kaplan@jhu.edu}
\affiliation{Department of Physics and Astronomy, Johns Hopkins University, Baltimore, MD 21218, USA}

\author{Surjeet Rajendran}
\email{srajend4@jhu.edu}
\affiliation{Department of Physics and Astronomy, Johns Hopkins University, Baltimore, MD 21218, USA}

\author{Harikrishnan Ramani} 
\email{hramani@stanford.edu}
\affiliation{Stanford Institute for Theoretical Physics, Stanford University, Stanford, CA 94305, USA}

\author{Erwin H. Tanin}
\email{etanin1@jhu.edu}
\affiliation{Department of Physics and Astronomy, Johns Hopkins University, Baltimore, MD 21218, USA}

\preprint{UMD-PP-022-05}

\begin{abstract}
We study the solar emission of light dark sector particles that self-interact strongly enough to self-thermalize. The resulting outflow behaves like a fluid which accelerates under its own thermal pressure to highly relativistic bulk velocities in the solar system. Compared to the ordinary non-interacting scenario, the local outflow has at least $\sim 10^3$ higher number density and correspondingly at least $\sim 10^3$ lower average energy per particle. We show how this generic phenomenon arises in a dark sector comprised of millicharged particles strongly self-interacting via a dark photon. The millicharged plasma wind emerging in this model has novel yet predictive signatures that encourages new experimental directions. This phenomenon demonstrates how a small step away from the simplest models can lead to radically different outcomes and thus motivates a broader search for dark sector particles.
\end{abstract}

\maketitle

Light particles with some coupling to the Standard Model (SM) can be produced in the Sun with $\sim \text{keV}$ energies. The luminosity of such particles is strongly limited by stellar cooling arguments \cite{Raffelt:1990yz}. If these particles simply free stream away as soon as they are produced in the Sun, 
the outcome is an outflow of $\sim \text{keV}$ energy particles whose particle-number flux is currently too low to be detected near the Earth. However, interactions within the dark sector are in general poorly constrained, and, as we will show, the story can change dramatically if one takes them into account.

We focus on the predictive scenario where the interactions within the dark sector allow these particles to locally thermalize via number-changing processes. There are two natural outcomes of this scenario. (1) While the dark-particle luminosity is still limited by cooling bounds, the self-thermalization of these particles has the effect of enhancing the resulting particle-number flux at the expense of lowering the average energy per particle.  Furthermore, (2) once local thermal equilibrium can be established, the mean free path of these particles can be microscopically small and on macroscopic scales they collectively display hydrodynamic behaviour which further modifies the properties of the outflow. The actual dynamics is a mixture of both effects and this gives rise to novel experimental and astrophysical signatures.

As long as these particles are relativistic, their thermal pressure will continually convert thermal energy into bulk fluid motion. In a way mathematically analogous to the Parker solar wind model \cite{parker1965dynamical}, this eventually leads to a steady outflow of the fast dark-particle fluid, which we refer to as the \textit{dark solar wind}.

We elaborate these points in the remainder of this {\it Letter}. For concreteness, we adopt a model of dark fermions interacting via dark photons as a representative of a self-interacting dark sector. In order to have this sector produced in the Sun, the dark photon is assumed to have a small kinetic mixing with the SM photon \cite{Holdom:1985ag}, thus making the dark fermions effectively millicharged. This property also enables this dark solar wind to be detectable on Earth.

\paragraph*{\textbf{Model.---}}
Dark sector particles need to satisfy the following conditions to form a dark solar wind:
\begin{enumerate}
    \item  There exist a coupling between the dark sector particles and SM particles so that they are produced in the Sun, and this coupling must be weak enough to avoid the stellar cooling bound.
    
    \item Their self-interaction is sufficiently strong to allow them to self-thermalize after being produced. 
    
    \item They are light enough to remain relativistic even after self-thermalization.
\end{enumerate}
There are various dark sector models to meet these conditions, but in this paper we focus on a specific model, a dark photon and a dark fermion which becomes a millicharged particle after photon-dark photon mixing.

Our Lagrangian is described by 
\beq
\mathcal{L}_D &=& - \frac{1}{4} F'_{\mu \nu} F'^{\mu \nu} - \frac{\epsilon}{2} F'_{\mu \nu} F^{\mu \nu} 
\nonumber\\&& + \, \, \bar \chi \pL i \gamma^\mu \partial_\mu + g_D \gamma^\mu A'_\mu -m_\chi  \pR \chi\,, 
\label{model}
\eea
where $\chi$ is a dark fermion with mass $m_\chi$, $A'$ is the massless dark photon, $\epsilon$ is the mixing angle between the SM photon and the dark photon, and $g_D=\sqrt{4 \pi \alpha_D}$ is the dark gauge coupling. Once the kinetic terms are diagonalized, $\chi$ is effectively millicharged and couples to the SM photon with electric charge $\epsilon g_D$.
This model has been studied in various contexts (see e.g. \cite{Alexander:2016aln,Battaglieri:2017aum} for reviews), including the recent works  \cite{Arvanitaki:2021qlj, Dvorkin:2019zdi,Mathur:2020aqv}.

In the main part of the analysis, we limit ourselves to the parameter space where $m_\chi$ is light enough that the resulting phenomenology is equivalent to that of  massless fermions. We will clarify the boundary of this regime and comment on how the phenomenology would change for heavier $m_\chi$.

\paragraph{\textbf{Production in the Sun.---}}
Due to their non-zero electric charge, $\chi$ particles can be pair created from the hot and dense SM plasma inside the Sun. For a dark fermion $\chi$ with mass much smaller than the solar temperature ($m_\chi \ll T_\odot$), the dominant production mechanism is via transverse plasmon decays. The number-density production rate $\dot{n}$ and the power per unit volume $\dot{Q}$ in the form of $\chi\bar{\chi}$ pairs through this channel were found analytically in \cite{Chang:2019xva}. The calculation details are shown in the Supplemental Material. To obtain their numerical values, we pair this analysis with the solar temperature and density profiles of \cite{Bahcall:2004pz}. For the crude estimates in this \textit{Letter}, we will use the following values of $\dot{n}$ at the center of the Sun and the $\chi$ production luminosity 
\bea
\dot{n}_c &\equiv& \, \dot{n}(r=0)\nonumber\\ 
&\sim & \, 2 \times 10^{9} \left(\frac{\epsilon}{10^{-15}}\right)^2\left(\frac{\alpha_D}{1}\right) \, \textrm{cm}^{-3}\textrm{s}^{-1} \, ,\\
L_\chi &\equiv& \, {\displaystyle \int_{V_\odot}} \dot{Q} dV \nonumber\\ 
&\sim & \, 8 \times 10^{36} \left(\frac{\epsilon}{10^{-15}}\right)^2\left(\frac{\alpha_D}{1}\right) \, \MeV \, \textrm{s}^{-1} \, .
\eeq
If the $\chi$ electric charge $\epsilon g_D$ is too large, the $\chi$ particles produced in stars carry away anomalously large amounts of energy thereby changing the evolutionary history of the stars. Bounds from the non-observation of such an anomalous evolution in red giants sets the most stringent stellar bound on the $\chi$ electric charge:  \cite{Vogel:2013raa}
\begin{equation}
    \epsilon\alpha_D^{1/2}\lesssim 2 \times 10^{-15} \label{coolinglimit}
\end{equation}

\paragraph*{\textbf{Self-thermalization.---}}
The initial population of $\chi$ particles produced in the Sun is in a state far from (local) thermal equilibrium. Here, we work out a sufficient condition for these particles to achieve thermalization. 

The newly pair-created $\chi$ particles from plasmon decays in the Sun have a ``hard" energy spectrum, with a typical energy $E_{\rm hard}$ roughly given by the temperature at the core of the Sun ($T_\odot$)
\begin{equation}
    E_{\rm hard}\sim T_\odot \sim 1\text{ keV} \, .
\end{equation}
Conservatively we start with the lowest possible abundance of these hard particles arising from the free-streaming regime. Since these particles are produced relativistically, they typically stay inside the \textit{solar core radius} $r_{\rm core}$, which we take as  $r_{\rm core}\approx 0.2 r_\odot$\footnote{Note that the SM plasma, and hence $\dot{n}$, is roughly uniform within the Sun's core radius $r_{\rm core}\approx 0.2 r_\odot$ and decays rapidly outside of it. For this reason it is more appropriate to use $r_{\rm core}$ instead of the full solar radius $r_\odot$ in our estimates.}, for a period of $\sim r_{\rm core}$. Hence, their starting number density is
\bea
n_{\rm hard} &\sim& \dot{n}_c r_{\rm core} \nonumber\\
&\sim& 8 \times 10^{8} \left(\frac{\epsilon}{10^{-15}}\right)^2\left(\frac{\alpha_D}{1}\right) \text{ cm}^{-3}
\eea
These $E_{\rm hard}$ and $n_{\rm hard}\ll E_{\rm hard}^3$ are a much higher average energy and a much lower number density compared to their would-be thermal equilibrium values for the same energy density. In order to thermalize, these particles must decrease their average energy and increase their number density.

The high-energy, under-occupied initial state of the $\chi$ particles produced in the Sun in our scenario resembles that of the products of perturbative inflaton decay in the early universe, a well-studied scenario \cite{Davidson:2000er,Harigaya:2013vwa,Mukaida:2015ria}. The subsequent thermalization of such particles proceeds dominantly through inelastic processes. While kinematics  forbids a single $\chi$ particle from spontaneously emitting a dark photon $\gamma_D$, particle production instead proceeds through soft bremsstrahlung. The rate of such a $2\rightarrow 3$ process is roughly the rate of the enabling soft scattering process $\Gamma_{2\rightarrow 2}^{\rm soft}$ multiplied by a factor of $\alpha_D$ for the $\gamma_D$ emission. Despite the extra $\alpha_D$ suppression, the rate $\Gamma_{2\rightarrow 3}$ is enhanced due to the fact that it is dominated by soft momentum exchanges, whose cross-section is large. This rate $\Gamma_{2\rightarrow 3}$ is also limited by the formation time of the emitted $\gamma_D$, i.e. the so-called Landau-Pomeranchuk-Migdal (LPM) effect \cite{Arnold:2001ba, Kurkela:2011ti}, but even so it is still much faster than other effects, such as the large-angle elastic scattering rate $\Gamma_{2\rightarrow 2}^{\Delta\theta \sim 1}$.

The soft $2\rightarrow 2$ scattering rate is infrared divergent and given by
\begin{equation}
    \Gamma_{2\rightarrow 2}^{\rm soft}\sim\frac{\alpha_D^2n_{\rm hard}}{k_{\rm min}^2}
\end{equation}
where $k_{\rm min}$ is the IR cutoff of the dark photon momentum, $k>k_{\rm min}$. In our setup, $k_{\rm min}$ is set by the pre-thermalization dark-sector Debye scale $\omega_D^{\rm pre}$, which is initially
\bea
    k_{\rm min} \sim \omega_D^{\rm pre} &\sim& \left(\frac{\alpha_D n_{\rm hard}}{E_{\rm hard}}\right)^{1/2}\nonumber\\
    &\sim& 7 \times 10^{-5} \left(\frac{\epsilon}{10^{-15}}\right)\left(\frac{\alpha_D}{1}\right) \, \eV \label{kmindebye}
\eea
A $\chi$ particle with an incoming momentum $p_{\rm in}\sim E_{\rm hard}$ can scatter with another $\chi$ particle, become off-shell, and emit an extra $\gamma_D$ with a momentum $k\lesssim p_{\rm in}$. The rate for such a $2\rightarrow 3$ process is given by \cite{Garny:2018grs}
\bea\label{23}
    \Gamma_{2\rightarrow 3} &\sim& \alpha_D\text{ min}\left(\Gamma_{2\rightarrow 2}^{\rm soft},t_{\rm form}^{-1}\right)\nonumber\\
    &\sim&  \alpha_D\Gamma_{2\rightarrow 2}^{\rm soft}\text{min}\left(1,\sqrt{\frac{k}{k_{\rm LPM}}}\right)
\eea
where $t_{\rm form}^{-1}$ sets an upper bound on the splitting rate due to the fact that only one dark photon can be emitted in the timescale $t_{\rm form}$ it takes to resolve the dark photon. This leads to a suppression of the splitting rate known as the LPM effect. In our case, $t_{\rm form}\sim \sqrt{E_{\rm hard}^2/\alpha_D^2n_{\rm hard}k}$ \cite{Garny:2018grs} and on the second line we re-expressed it in terms of the threshold momentum of the emitted $\gamma_D$ below which the LPM effect becomes relevant
\begin{equation}
    k_{\rm LPM}\sim \frac{E_{\rm hard}^2\Gamma_{2\rightarrow 2}^2}{\alpha_D^2 n_{\rm hard}}\sim\alpha_D\left(\frac{E_{\rm hard}}{k_{\rm min}}\right)^2E_{\rm hard}\gg E_{\rm hard}
\end{equation}
The finding that $k_{\rm LPM}\gg E_{\rm hard}$ tells us that the LPM effect is always important in our setup and the $2\rightarrow 3$ process is the fastest for $k\sim E_{\rm hard}$, which boils down to the rate being $\Gamma_{2\rightarrow 3}\sim \alpha_D^{3/2}\omega_D^{\rm pre}$. In the subsequent $2\rightarrow 3$ splittings, $E_{\rm hard}$ goes down, $n_{\rm hard}$ goes up, $k_{\rm min}$ goes up, which means $\Gamma_{2\rightarrow 3}$ will keep increasing. Through processes such as $\gamma_D\chi\rightarrow\bar{\chi}\chi\chi$, the abundance of $\chi$ pairs increases with the abundance of the dark photons $\gamma_D$ leading to an acceleration to thermal equilibrium in the dark sector. Hence, the bottleneck lies in the beginning and the requirement for achieving thermalization with this process is $\Gamma_{2\rightarrow 3}^{\rm initial}r_{\rm core}\gtrsim 1$, or
\begin{equation}
\epsilon \alpha_D^{5/2} \gtrsim 2 \times 10^{-26}
\label{thermalizationcondition}
\end{equation}

We have identified a sufficient condition for the dark sector particles to thermalize via bremsstrahlung processes. Other inelastic processes not considered here are expected to bring the dark sector phase-space distribution closer to thermalization. We have checked that elastic processes such as scatterings within the dark sector, scatterings with SM particles, and the bending of trajectories due to the solar magnetic field are negligible in the pre-thermalization phase. Even if these effects were not weak they would only help trap the $\chi$ particles, making it even easier for them to thermalize.

\paragraph*{\textbf{Dark Solar Wind.---}}
If the dark sector particles manage to completely self-thermalize, their mean free path would generically be far smaller than the length scales that dictate their collective macroscopic dynamics\footnote{This is trivially satisfied for the model \eqref{model} we are considering if the condition \eqref{thermalizationcondition} is met. More specifically, we have checked a posteriori that the mean free time of the dark particles as seen in the Sun's frame is easily much smaller than the timescale for, e.g., the density of particles to change by an $O(1)$ factor due to the fluid expansion.}. In that limit, such particles behave like a perfect fluid whose properties can be described in a largely model-independent way. Given the billion-year age of the Sun, it is likely that this fluid has relaxed by now to a steady state, described by the following time-independent energy and momentum equations\footnote{Numerical simulations \cite{1978ApJ...221..661S, Keto2020StabilityAS} of analogous systems \cite{parker1965dynamical, 1979A&A....73..171G, 1980MNRAS.191..571P, 1992ApJ...384..587T,2002PhRvE..66f6303R} suggest that the steady state solution is an attractor and the relevant relaxation timescale is given by the hydrodynamical timescale, which for our solar system is roughly the flow time to 1 AU, namely $\sim 10$ minutes.}
\bea
    \frac{1}{r^2}\partial_r\left[r^2\gamma^2v\left(\tilde{\rho}+\tilde{p}\right)\right]&=&\dot{Q} \label{energyeq}\\
    \frac{1}{r^2}\partial_r\left[r^2\gamma^2v^2\left(\tilde{\rho}+\tilde{p}\right)\right] &=&-\partial_r\tilde{p} \label{momentumeq}
\eea
where $r$ is the radial position with respect to the center of the Sun, $\gamma=(1-v^2)^{-1/2}$ is Lorentz factor associated to the radial bulk velocity $v$ of the fluid, $\tilde{\rho}(r)$ and $\tilde{p}(r)$ are the comoving density and pressure of the fluid, and $\dot{Q}(r)$ is the power per unit volume injection from the Sun in the form of $\chi\bar{\chi}$ pairs. Quantities with a tilde $\tilde{}$ on top of it are defined in the rest frame of the fluid and those without it are defined in Sun's frame.

For simplicity, we assume here that the $\chi$ particles are massless or sufficiently light that they are adequately described as a radiation-dominated fluid with $\tilde{\rho}=3\tilde{p}=a\tilde{T}^4$, with $a$ a constant. Integrating the energy equation \eqref{energyeq} gives us the comoving temperature $\tilde{T}$ of the fluid in terms of the fluid velocity $v$
\begin{equation}
    \frac{4a}{3}\tilde{T}^4=\frac{\int_0^r \dot{Q}(r') 4\pi r^{\prime 2}dr^\prime }{4\pi \gamma^2v r^2}\label{Tvrelation}
\end{equation}
Substituting the above into the momentum equation \eqref{momentumeq} results in an equation for the fluid velocity only
\begin{equation}
    \left(\frac{1/3-v^2}{1/3+v^2}\right)\frac{\partial \ln v}{\partial \ln r}=f(r)-\frac{2(1-v^2)}{1+3v^2}\label{veq}
\end{equation}
where we defined a source function
\begin{equation}
    f(r)\equiv \frac{\dot{Q}(r)r^3}{\int_0^r \dot{Q}(r')r'^2dr'}
\end{equation}
The first term on the RHS of \eqref{veq} is due to the energy injection ({\it i.e.} inertia injection) from the Sun, while the second term stems from the pressure gradient of the fluid. Since the source $\dot{Q}(r)$ enters only via the dimensionless quantity $f(r)$, the resulting velocity profile $v(r)$ does not depend on the normalization of $\dot{Q}$, but only on the radial variation of the quantity $f(r)$. In particular, for the model \eqref{model} under consideration $f(r)$ is independent of both $\epsilon$ and $\alpha_D$.

\begin{figure}
    \centering
    \includegraphics[width=\linewidth]{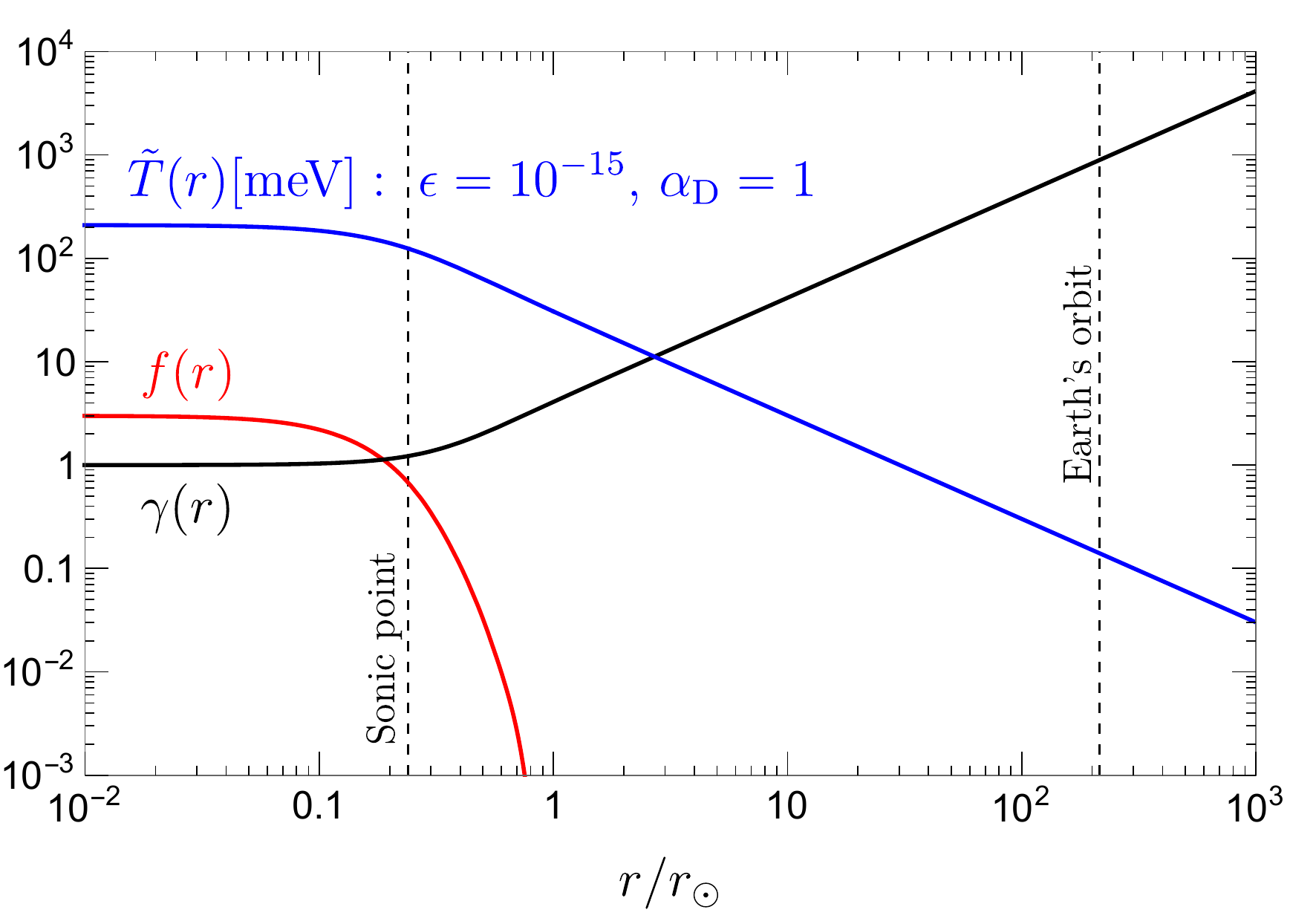}
    \caption{Dark solar wind profiles as a function of the radial distance $r$ from the Sun. The Lorentz factor $\gamma(r)$ solves the fluid velocity equation \eqref{veq} with the transonic boundary condition \eqref{boundaryconditions}. The source function $f(r)\equiv \dot{Q}(r)r^3/\int_0^r \dot{Q}(r')r^{\prime 2}dr'$ is computed with the millicharged-particle energy density injection rate $\dot{Q}$ of \cite{Chang:2019xva} and the numerical solar profiles of \cite{Bahcall:2004pz}. Both $\gamma(r)$ and $f(r)$ are independent of $\epsilon$ and $\alpha_D$. The comoving temperature $\tilde{T}(r) \propto (\epsilon^2 \alpha_D)^{1/4}$ of the fluid is evaluated from the integrated energy equation \eqref{Tvrelation} for $\epsilon= 10^{-15}$ and $\alpha_D=1$. Also shown are the location of the sonic point, $r=0.24r_\odot$, and the Earth's orbit radius, $r=215r_\odot$.}

    \label{fig:fluidprofiles}
\end{figure}

In order to solve \eqref{veq} we need to specify some boundary conditions. Since the Sun produces $\chi$ particles with no net radial momenta to begin with, $v$ must vanish at the origin. As we go to larger radii, there are two possible types of profiles, depending on whether or not $v$ goes above the speed of sound $c_s=1/\sqrt{3}$ during the outflow. Subsonic solutions\footnote{This includes hydrostatic solutions.}, where $v<c_s$ all the way, predict $v\propto r^{-2}$ at large distances from the Sun which, in turn, implies a non-zero comoving temperature $\tilde{T}$ at infinity through \eqref{Tvrelation}. However, the latter boundary condition is unphysical. Unless we add new ingredients to the model that provide pressure support on the fluid far from the Sun, e.g. through the high cosmic abundance of some particles interacting with the fluid, we expect $\tilde{T}$ to vanish at large $r$. This leaves us with the remaining possibility, namely the transonic solution, in which case the fluid velocity increases from subsonic speeds ($v<c_s$) at small $r$ through the \textit{sonic point} ($v=c_s$) to supersonic speeds ($v>c_s$) at large $r$. The monotonically increasing velocity implies a monotonically decreasing $\tilde{T}$, thus satisfying the vanishing $\tilde{T}$ boundary condition at infinity. 

The sonic point ($v=1/\sqrt{3}$) can be crossed smoothly only if it coincides with the zero of the RHS in \eqref{veq} for $v=1/\sqrt{3}$. The latter most likely occurs at around the Sun's core radius $r_{\rm core}$, where $\dot{Q}$ starts to drop rapidly, and numerically we found it to lie at $r=r_{\rm sonic}\approx 0.24 r_\odot$. Thus, we require
\begin{equation}
    v=\frac{1}{\sqrt{3}}\quad\text{at}\quad r=r_{\rm sonic}\approx 0.24r_\odot \label{boundaryconditions}
\end{equation}
This completes the boundary conditions for numerically solving the velocity equation \eqref{veq}. We plot the resulting transonic Lorentz factor profile $\gamma(r)$, together with the source function $f(r)$, and the comoving fluid temperature $\tilde{T}$ from \eqref{Tvrelation} in Fig.~\ref{fig:fluidprofiles}. It shows that beyond the sonic point $r\gtrsim r_{\rm sonic}$ the Lorentz factor $\gamma$ of the fluid flow asymptotes toward the well-known fireball solution \cite{Piran:1993jm}, $\gamma\sim r/r_{\rm sonic}$, for an adiabatically expanding fluid. In fact, as shown in Ref.~\cite{Piran:1993jm}, when $\gamma\gg 1$ the fireball solution $\gamma\sim r/r_{\rm sonic}$ solves not only the the sourceless $\dot{Q}=0$ steady-state fluid equations \eqref{energyeq} and \eqref{momentumeq} but also the \textit{time-dependent} fluid equations. Thus, we expect $\gamma\sim r/r_{\rm sonic}$ to hold robustly outside the Sun even if for some reason the fluid flow deviates from the assumed steady-state solution inside the Sun.

The structure of the fluid equation \eqref{veq} considered here and the singling out of the transonic solution are mathematically analogous to that of Parker's solar wind \cite{parker1965dynamical} (see also Bondi accretion \cite{1952MNRAS.112..195B}). However, the physical mechanisms behind them are completely different. Parker's solar wind is isothermal, non-relativistic, and accelerated by an interplay between pressure gradient and gravity. On the other hand, the dark solar wind is adiabatic, relativistic, and accelerated by an interplay between the pressure gradient and energy/inertia injection from the Sun.

\paragraph*{\textbf{Properties near the Earth.---}}

As the flow expands to larger $r$ and accelerates to higher Lorentz factors $\gamma$, the comoving temperature of the fluid $\tilde{T}$ cools down adiabatically according to \eqref{Tvrelation}. The integral in \eqref{Tvrelation} for $r\gtrsim r_{\rm core}$ yields the luminosity $L_\chi$ of the $\chi$ particles produced in the Sun, resulting in $(4a/3)\tilde{T}^4\approx \gamma^{-2}(L_\chi/4\pi r^2)$ for $v\approx 1$. Interestingly, in the highly-relativistic limit $\gamma\gg 1$ expected at $r\gg r_{\rm sonic}$ the scalings with $r$ of the average energy per particle $\left<E\right>\sim \gamma \tilde{T}\approx \text{const.}$ and number density $n\sim \gamma \tilde{T}^3\propto r^{-2}$ in the Sun's frame are identical to those in the free streaming case, i.e. it is as if these particles simply free streamed from the surface at which the fireball approximation starts to hold ($r\sim r_{\rm sonic}$). The latter is understandable because the acceleration of the fluid to relativistic bulk velocities manifests itself at the particle level as the velocities of the particles becoming increasingly radial the farther they are from the Sun (relativistic beaming).

All things considered, the average energy $\left<E\right>$ and number density $n$ of the dark particles at $r\gg r_{\rm sonic}$ in the Sun's frame are given up to $O(1)$ factors by
\begin{align}
    \left<E\right>&\sim  \left(\frac{L_\chi}{ r_{\rm sonic}^2}\right)^{1/4}\sim 1\text{ eV}\left(\frac{L_\chi}{10^{-2}L_\odot}\right)^{1/4}\\
    n&\sim \frac{n_{\rm FS}\left<E\right>_{\rm FS}}{\left<E\right>}\sim 10^3n_{\rm FS}\left(\frac{L_\chi}{10^{-2}L_\odot}\right)^{-1/4}
\end{align}
By contrast, in the free-streaming case the average energy per particle is given by the core temperature of the Sun $\left<E\right>_{\rm FS}\sim \text{keV}$ and if these particles are massless energy conservation then gives $n_{\rm FS}\sim L_\chi/r^2\left<E\right>_{\rm FS}$. When the dark particle luminosity saturates the cooling limit, $L_\chi\sim 10^{-2} L_\odot$, the dark solar wind gives $\sim 10^{3}$ lower average energy $\left<E\right>$ and $\sim 10^{3}$ higher number density $n_\chi$ compared to those in the free-streaming case. The results in the two cases deviate even more for $L_\chi\ll 10^{-2}L_\odot$. For the dark fermion dark photon model \eqref{model} considered in this {\it Letter}, and with the $O(1)$ factors included, the results at $r\gg r_{\rm sonic}$ are
\begin{align}
    \gamma&\approx 893 \left(\frac{r}{1\text{ AU}}\right)\\
    \tilde{T}&   \approx   0.14\, \meV\left(\frac{\epsilon}{10^{-15}}\right)^{\frac{1}{2}}\left(\frac{\alpha_D}{1}\right)^{\frac{1}{4}}\left(\frac{1\text{ AU}}{r}\right) \\
    \langle E \rangle&\approx 4 \gamma \tilde{T} \approx 0.5\, \eV \left(\frac{\epsilon}{10^{-15}}\right)^{\frac{1}{2}}\left(\frac{\alpha_D}{1}\right)^{\frac{1}{4}} \label{avgenergy}\\
    n &= \frac{5 \zeta(3)}{\pi^2}\gamma \tilde{T}^3\approx \frac{2 \times 10^{5}}{\text{cm}^3} \left(\frac{\epsilon}{10^{-15}}\right)^{\frac{3}{2}}\left(\frac{\alpha_D}{1}\right)^{\frac{3}{4}}\left(\frac{1\text{ AU}}{r}\right)^2 \label{numdensity}    
\end{align}
where we have used $\tilde{\rho}=a\tilde{T}^4$ with $a=(2+4\times 7/8)(\pi^2/30)$ corresponding to dark photons and $\chi\bar\chi$ pairs in obtaining the above results, and $\left<E\right>$ was found by averaging over all species using their spectra as seen on Earth. Naively, the relevant spectrum of these particles on Earth would be that of a thermal distribution boosted by $\gamma_{\text{AU}}\approx 893$. The spectrum that is actually seen in an Earth-based detector is however biased by the fact that the detector samples dark particles in an anisotropic way, e.g. more from the side facing the incoming dark flow than from the opposite side. Accounting for both these effects yields a distribution similar to that of a $\gamma_{\text{AU}}^{-2}$ rescaled thermal distribution \cite{PhysRev.176.1451}
\begin{equation}
    \frac{dn}{dE d\Omega}=\frac{1}{\gamma_{\text{AU}}^2}\frac{g_*}{(2\pi)^3}\frac{E^2}{e^{E/T_{\rm eff}}\pm 1} \label{boosteddist}
\end{equation}
with the degrees of freedom of the particle of interest $g_*$ and a direction-dependent effective temperature
\begin{equation}
    T_{\rm eff}(\theta)\approx \frac{\tilde{T}}{\gamma_{\text{AU}}(1-v_{\rm AU}\cos\theta)}
\end{equation}
where $v_{\rm AU}=(1-1/\gamma_{\rm AU}^2)^{1/2}$, $d\Omega$ is a differential solid angle, and $\theta$ is the angle of the particle velocity relative to the local direction of the dark solar wind. Due to relativistic beaming the angular distribution \label{boosteddist} has a strong support only within $\theta\lesssim \gamma_{\rm AU}^{-1}$.

\begin{figure}[t]
\centering
\includegraphics[width=\linewidth]{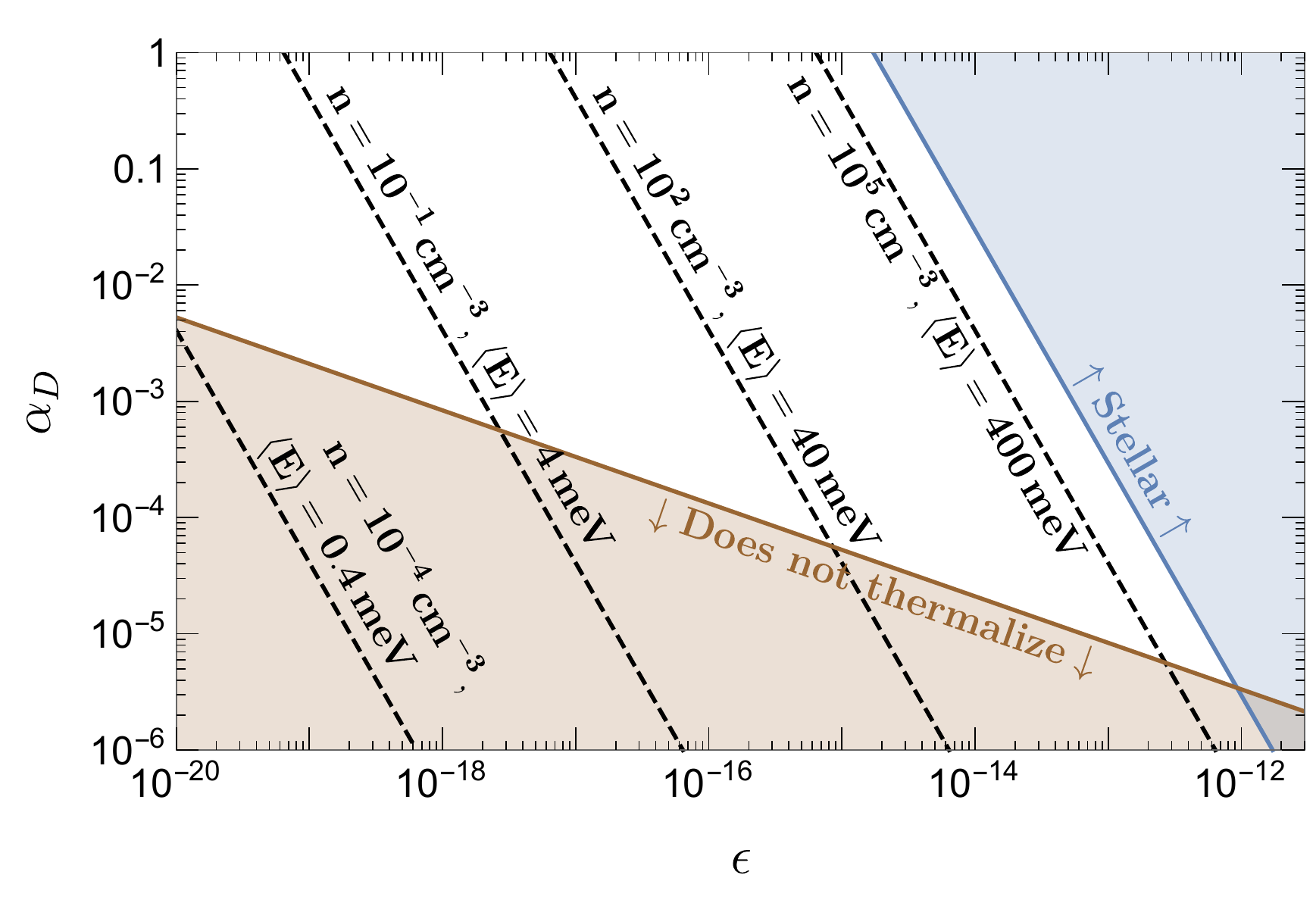} 
\caption{Viable parameter space and predictions. The blue region violates the stellar cooling limit \eqref{coolinglimit}. The red region does not satisfy the thermalization requirement \eqref{thermalizationcondition}. The dashed lines correspond to different combinations of dark solar wind density $n$ \eqref{numdensity} and average energy per particle $\left<E\right>$ \eqref{avgenergy} near the Earth.}
    \label{paramspace}
\end{figure}

\paragraph*{\textbf{Detection.---}}
We next briefly discuss detection prospects of dark solar wind on Earth. In principle, the dark photons as well as the dark fermions in the dark plasma can be detected. Since the prospects are futuristic, we only provide order-of-magnitude estimates and defer a systematic study for future work.

We start by considering experiments proposed for detecting non-relativistic dark photons making up the dark matter via absorption \cite{Hochberg:2016sqx,Bunting:2017net,Knapen:2021bwg,Mitridate:2021ctr}. A subset of these experiments are also sensitive to the relativistic dark photons that have energies within the sensitive ranges of the detectors. However, the absorption rate for the relativistic dark photons are suppressed compared to their non-relativistic dark matter counterparts due to inferior number density as well as suppression of the form $\left(\alpha_D^{1/2} \tilde{T}/\langle E\rangle\right)^\beta$, where $\beta=4(2)$ for the transverse (longitudinal) relativistic dark photons. The latter comes from taking the relativistic limits of the dark-photon absorption rates in \cite{An:2014twa}. Due to this suppression, our preliminary estimate suggests that the sensitivity of recently proposed dark-photon-absorption experiments to dark solar wind is still a few orders of magnitude above the cooling bound.

The dark fermions in the dark solar wind can potentially be probed through their elastic or inelastic scattering with electrons in dark matter direct detection experiments. In elastic scattering, since the dark fermions have energies much lower than the mass of the electron, the energy deposited in the electron is suppressed. The maximum average energies for the dark fermions in this scenario are in the $\sim$ eV range resulting in energy depositions $\lessapprox$ meV. This is lower than the planned thresholds of near-term experiments but could potentially be detected in the future. Inelastic processes wherein a bound electron is kicked out of its shell can kinematically permit the dark fermion to lose $\mathcal{O}\left(1\right)$ of its kinetic energy {\it i.e.} almost $\sim$ eV. While these processes are kinematically more favorable, the cross-section for such inelastic processes is suppressed by the momentum transferred during the process, requiring detectors with larger target masses to probe phenomenologically interesting parts of parameter space. We discuss these possibilities in detail in future work.

The dark plasma flow studied here has properties that are very distinct from that of cold dark matter, permitting detection strategies beyond scattering in direct detection experiments. The uni-directional, relativistic and strongly-coupled nature of the flow may allow us to probe this parameter space through experiments in the same spririt as the so-called \textit{direct deflection} \cite{Berlin:2019uco,Berlin:2021kcm}. The idea is to somehow perturb the dark plasma wind and measure its back-reaction in the form of SM electric or magnetic fields downstream. The generically tiny dark Debye length of the dark plasma in our scenario quickly erases any static dark electric field in the dark plasma once it is removed from the perturber. However, dark electric currents may persist long enough to be detected \cite{Baym:1995fk}. It would be interesting to quantify this nontrivial dynamics with the help of numerical simulations.

\paragraph*{\textbf{Discussion.---}}

We pointed out a new generic phenomenon, here referred to as \textit{dark solar wind}, that arises in a generic light dark sector with sufficiently weak SM interactions to avoid stellar cooling limits and sufficiently strong, number-changing self-interactions to self-thermalize upon emission. Unlike in the free-streaming scenario, the solar emission in this regime is less energetic, denser, and behaves like a relativistically expanding fluid. Since the properties of this fluid are dictated by thermal-equilibrium and steady-state hydrodynamics, they are not sensitive to the details of the underlying microphysics. We considered dark fermions charged under a dark photon that kinetically mixes with the SM photon as an example, spelled out a sufficient condition for achieving thermalization in the dark sector, numerically solved the hydrodynamic equations for the resulting fluid, and worked out their properties on Earth or elsewhere in the solar system. There might exist parameter space in other self-interacting dark sector models such as light non-Abelian sectors (motivated by potential solutions to the cosmological constant problem ~\cite{Graham:2017hfr, Graham:2019bfu, Berghaus:2020ekh, Ji:2021mvg}) and other models considered in \cite{Boddy:2014yra,March-Russell:2020nun} which also exhibits dark solar wind.

Though we assumed that the dark fermion $\chi$ is massless in our discussions, they are still valid for non-zero but light enough dark fermion mass $m_\chi$. The thermalization condition is unaffected as long as $m_\chi$ is less than the lowest relevant energy scale, namely the pre-thermalization debye frequency $\omega_D^{\rm pre}$, while the fluid dynamics is unchanged as long as the dark fermions remain relativistic, i.e. $m_\chi\ll \tilde{T}$, up to the radius of interest. As we increase $m_\chi$ from zero, we cross $\omega_D^{\rm pre}$ way before the comoving temperature near the Earth $\tilde{T}(r=1\text{ AU})$. We discuss this regime where the thermalization condition is parameterically different but the fluid dynamics and hence the model predictions are unchanged in the Supplemental Material.
We leave the explorations of yet higher $m_\chi$ regime as well as the effects of a non-zero dark photon mass for future work.

We assumed in our analysis that the transonic steady-state solution that solves the stationary fluid equation \eqref{veq} is stable and has been reached by now. We reiterate here that regardless of whether this stationary solution is reached and maintained, our predictions would still be $O(1)$ correct since the asymptotic fireball solution $\gamma\sim r/r_{\rm sonic}$ accurately solves the time-dependent fluid equation outside the Sun. That said, it would be interesting to run time-dependent hydrodynamical simulations in order to confirm that this is indeed true and also to study the potential time variability of the dark fluid properties.

The same SM couplings that produce the dark particles in the Sun could also lead to a relic cosmological abundance of these particles \cite{Chu:2011be,Evans:2019vxr,Dvorkin:2019zdi,Dvorkin:2020xga,Fernandez:2021iti}. Compared to the solar emissions, these cosmological relics are more model-dependent as they rely on the early universe cosmology and physics at higher energies. Nevertheless, we speculate on the impacts of such relics here. Firstly, limits on dark matter self-couplings from the Bullet cluster observation do not necessarily preclude the possibility that the same strongly-coupled elementary particles produced in the Sun also make up all the dark matter. Such astrophysical bounds can be relaxed or avoided if, for instance, the cosmological relics come in the form of clumpy gas \cite{Buckley:2017ttd} or composite bound states \cite{Cline:2013zca, Gresham:2017zqi}. Even if these relics only make up a small fraction of the dark matter, they can still have significant astrophysical impacts at galactic scales \cite{Fan:2013yva}. Furthermore, at the solar system scale, their interactions with the dark solar wind may lead to a spectrum of outcomes. On one extreme the dark matter may affect the dark solar wind by changing the boundary conditions of the fluid equations, while on the other extreme the dark solar wind may blow away and replace the typically assumed $\sim 0.3\text{ GeV}/\text{cm}^3$ dark matter energy density (or whatever its expected relic density is) near the Sun.

Analogous fluid-like emissions of dark particles may also appear in a wide range of astrophysical systems other than the Sun and be potentially detectable on Earth. The dark wind emanating from the Earth's core, while closer in distance to terrestrial detection, is inferior in number density and average energy in comparison to the dark solar wind due to far smaller dark luminosity. Other astrophysical objects are of course farther from us than the Sun, but their cores being much hotter and larger may yield new opportunities for probing dark sector parameter space. Furthermore, their emissions may differ from the solar one not only quantitatively but also qualitatively. For instance, supernovae are transient and inject momentum in addition to energy into the dark fluid, which might lead to completely different fluid solutions. This work is thus a first step toward exploring more general dark fluid outflows originating from a wide range of astrophysical systems.

\paragraph*{\textbf{Acknowledgments.---}}
We thank William DeRocco, Peizhi Du, Rouven Essig, Anson Hook, Gustavo Marques-Tavares, Mukul Sholapurkar, and Ken Van Tilburg for useful discussions. J.H.C is supported by the NSF grant PHY-1914731, the Maryland Center for Fundamental Physics, and the JHU Joint Postdoc Fund. D.E.K and S.R are supported in part by the U.S.~National Science Foundation (NSF) under Grant No.~PHY-1818899.  This work was supported by the U.S.~Department of Energy (DOE), Office of Science, National Quantum Information Science Research Centers, Superconducting Quantum Materials and Systems Center (SQMS) under contract No.~DE-AC02-07CH11359. 
S.R is also supported by the DOE under a QuantISED grant for MAGIS, and the Simons Investigator Award No.~827042. H.R
acknowledges the support from the Simons Investigator
Award 824870, DOE Grant DE-SC0012012, NSF Grant
PHY2014215, DOE HEP QuantISED award no. 100495,
and the Gordon and Betty Moore Foundation Grant
GBMF7946.

\bibliography{references}
\bibliographystyle{apsrev4-1}

\widetext


\begin{center}
\textbf{\large Supplemental Material: Dark Solar Wind}
\end{center}
\setcounter{equation}{0}
\setcounter{figure}{0}
\setcounter{table}{0}
\setcounter{page}{1}
\makeatletter
\renewcommand{\theequation}{S\arabic{equation}}
\renewcommand{\thefigure}{S\arabic{figure}}
\renewcommand{\bibnumfmt}[1]{[S#1]}
\renewcommand{\citenumfont}[1]{S#1}

\section{Plasma Effects in the Sun}
The dominant production channel of $\chi\bar{\chi}$ pairs in the Sun is through transverse plasmon decays, with the number density production rate $\dot{n}$ and energy density production rate $\dot{Q}$ given by \cite{Chang:2019xva}
\bea
\dot{n} &=& \int \frac{d^3 k}{(2 \pi)^3} g_{T} f_\textrm{BE} (\omega_{T}) \Gamma_{\gamma^*_T \rightarrow \chi \bar\chi} \\
\dot{Q} &=& \int \frac{d^3 k}{(2 \pi)^3} g_{T} f_\textrm{BE} (\omega_{T}) \omega_{T}  \Gamma_{\gamma^*_T \rightarrow \chi \bar\chi}
\, ,
\eea
where $g_{T}=2$ is the degree of freedom for transverse plasmon, $f_\textrm{BE}(E)$ is the Bose-Einstein distribution, $\omega_{T}(k)$ is the  energy of the transverse plasmon, and
\bea
\Gamma_{\gamma^*_T \rightarrow \chi \bar\chi} = \frac{\epsilon^2 \alpha_D}{3} \frac{\omega_{T}^2-k^2}{\omega_{T}} \, .
\eea
In the core of the Sun, the dispersion relation of the photon is modified by plasma effects. These effects are captured by the real parts of polarization tensors of the photon, $\text{Re}\Pi_{T,L}$, which come from the self-energy calculation with background electrons. In this work, we take the classical limit in \cite{Braaten:1993jw}, which gives
\bea
\text{Re}\Pi_T &=& \omega_p^2 \pL 1+\frac{k^2}{\omega_T^2}\frac{T}{m_e} \pR \, ,\\
\text{Re}\Pi_L &=& \omega_p^2 \pL 1+3\frac{k^2}{\omega_L^2}\frac{T}{m_e} \pR \, ,
\eea
for transverse and longitudinal plasmons with a 4-momentum $(\omega_{T,L},\vec{k})$ in a plasma with a temperature $T$. Here, $\omega_p$ is the plasma frequency,
\beq
\omega_p^2 = \frac{4\pi \alpha n_e}{m_e} \pL 1-\frac{5}{2} \frac{T}{m_e} \pR \, ,
\eeq
$m_e$ is the mass of the electron, and $n_e$ is the electron density in the plasma. Then the dispersion relation is given by
\bea
\omega_T^2 &=& k^2 + \text{Re}\Pi_T \, , \quad 0 \leq k < \infty \, , \\
\omega_L^2 &=& \text{Re}\Pi_L \, , \quad 0 \leq k < \omega_p \sqrt{1+\frac{3T}{m_e}} \, .
\eea
For the plasmon decay to dark fermions, the transverse mode dominates because the decay rate is proportional to the thermal mass, $\sqrt{\omega_{T,L}^2-k^2}$, and the longitudinal thermal mass drops quickly for large $k$. Also note we set the transverse vertex renormalization constant to 1 ($Z_T=1$) for our calculations.

\section{Dark Sector Thermalization for Massive Dark Fermions}

Though we assumed that $\chi$ is massless in deriving the thermalization requirement in the main text, it is still valid as long as the $\chi$ mass is below the pre-thermalization dark debye frequency, $m_\chi\lesssim \omega_D^{\rm pre}$. Here, we generalize the discussion to cases with more massive but still ultra-relativistic dark fermions $\chi$, while keeping the dark photons $\gamma_D$ massless. As described in the main text, the relaxation of the initially underpopulated $\chi$ particles produced in the Sun toward thermal equilibrium relies on plasma-induced radiative processes to increase the number of particles. Such processes have been studied extensively in the QED literature, see \cite{Arnold:2001ba,Peigne:2008wu}
for representative reviews on radiative loses in an abelian plasma and also \cite{Mukaida:2015ria, Garny:2018grs} which analyzed a setup similar to ours, namely the thermalization an underoccupied abelian plasma. The discussion below is an adaption of the results of these earlier works in our context.

\begin{figure}
    \centering
    \includegraphics[width=0.48\linewidth]{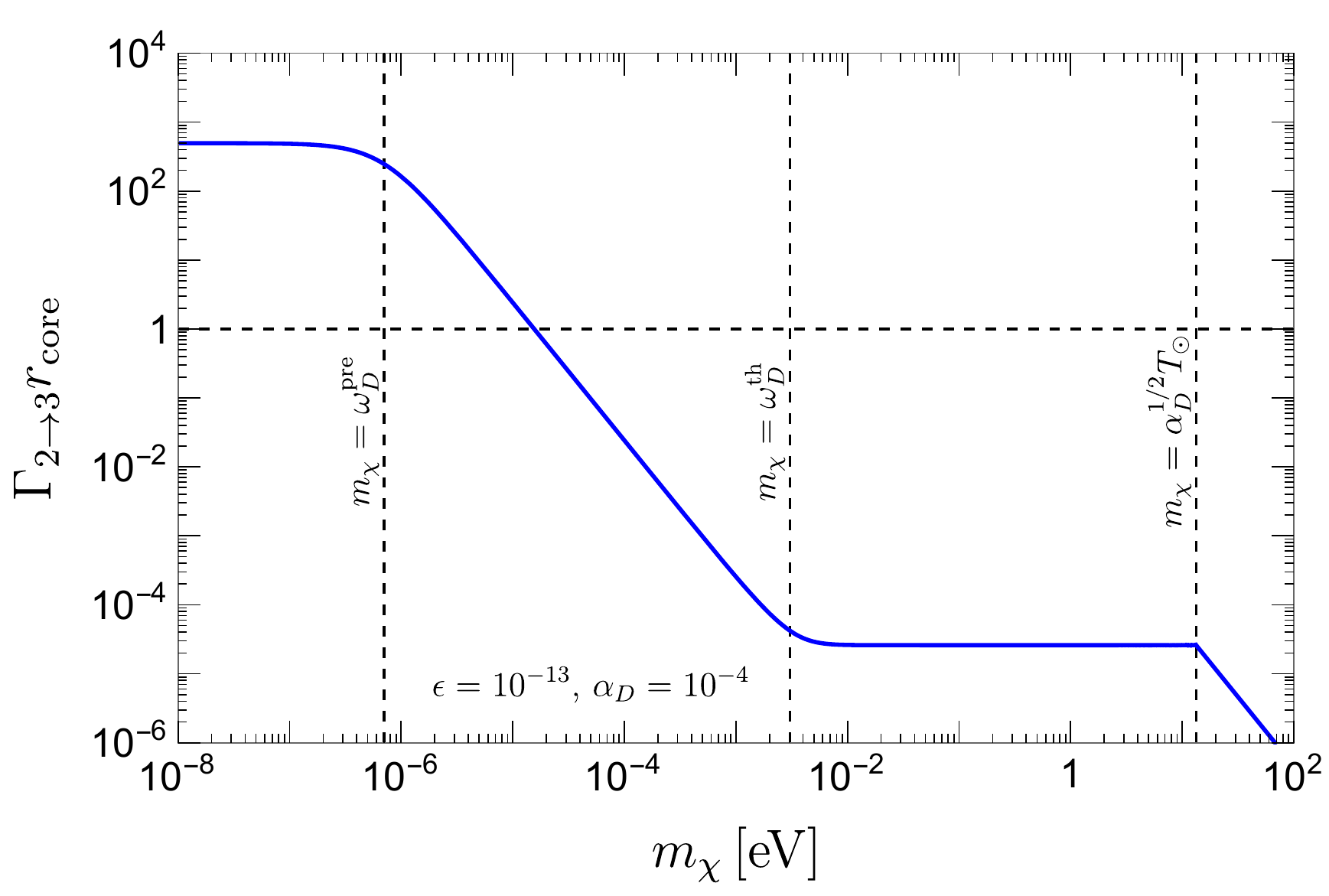}
    \includegraphics[width=0.48\linewidth]{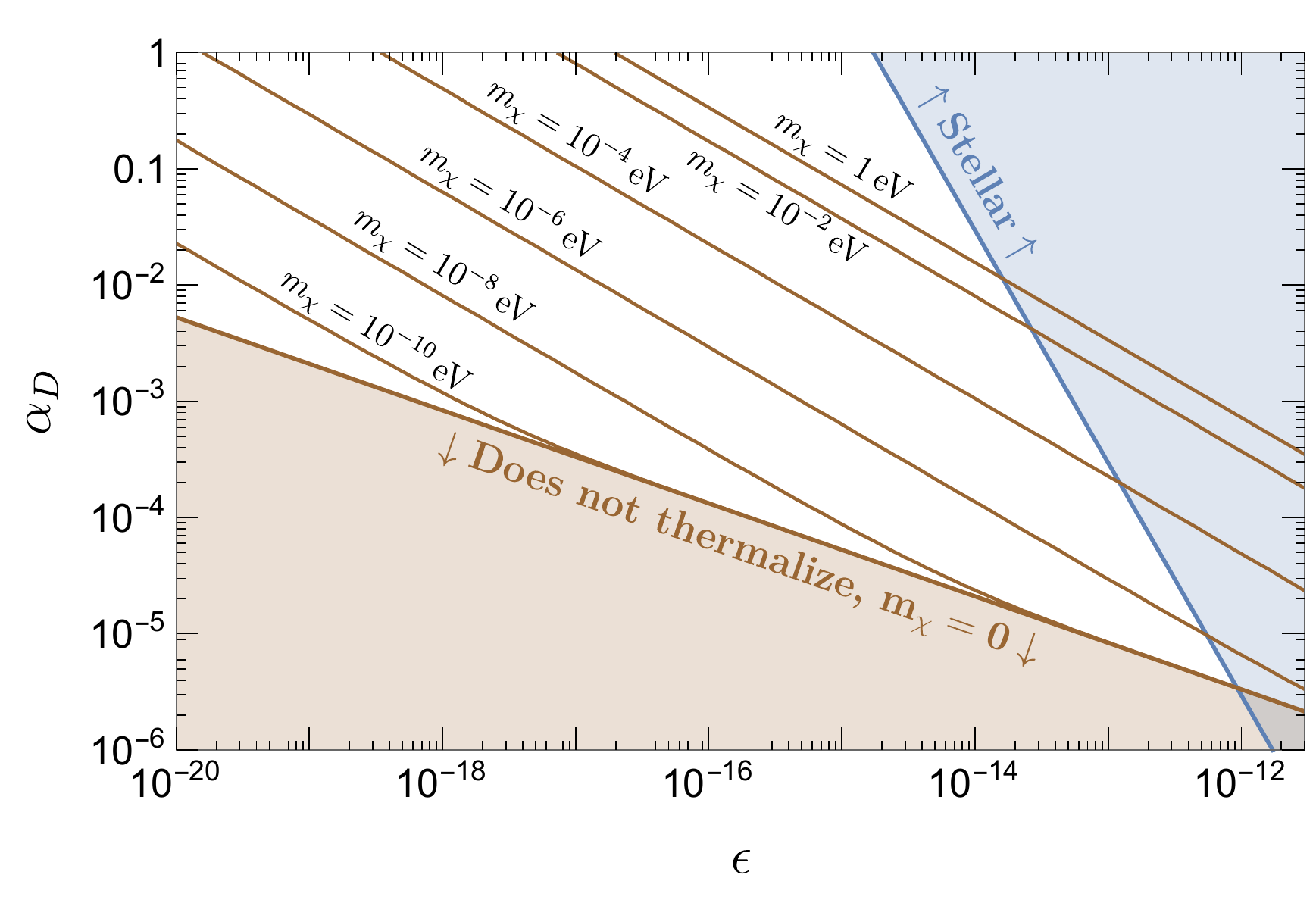}
    \caption{\textit{Left: }The pre-thermalization $\chi\chi\rightarrow\chi\chi\gamma_D$ splitting rate $\Gamma_{2\rightarrow 3}$ (up to $O(1)$ factors) as a function of the dark fermion mass $m_\chi$. This splitting process is efficient at thermalizing the dark sector if $\Gamma_{3\rightarrow 2}r_{\rm core}\gtrsim 1$, with $r_{\rm core}\approx 0.2r_\odot$. \textit{Right: }The viable parameter space for different values of $m_\chi$. Finite-mass effects lead to parameterically smaller $\Gamma_{2\rightarrow 3}$ which in turn makes the thermalization requirement more restrictive.}
    \label{fig:massivechi}
\end{figure}

As we increase $m_\chi$ from zero, different effects become relevant as it crosses the following key scales in the setup:
\begin{itemize}
    \item Pre-thermalization dark debye frequency inside the Sun
    \begin{equation}
        \omega_D^{\rm pre}\sim \sqrt{\frac{\alpha_D n_{\rm hard}}{E_{\rm hard}}}\sim 7\times 10^{-5}\left(\frac{\epsilon}{10^{-15}}\right)\left(\frac{\alpha_D}{1}\right)\text{ eV} \label{predebye}
    \end{equation}
    \item Post-thermalization dark debye frequency inside the Sun
    \begin{equation}
        \omega_D^{\rm th}\sim \alpha_D^{1/2}\tilde{T}\sim (\alpha_D^2n_{\rm hard}E_{\rm hard})^{1/4}\sim 0.3 \left(\frac{\epsilon}{10^{-15}}\right)^{1/2}\left(\frac{\alpha_D}{1}\right)^{3/4}\text{ eV}\label{postdebye}
    \end{equation}
    Note that we always have $\omega_D^{\rm th}\gg \omega_D^{\rm pre}$ because $n_{\rm hard}\ll \tilde{T}^3$ and $E_{\rm hard}\gg\tilde{T}$.
    \item The core temperature of the Sun $T_{\odot}\sim \text{ keV}$ multiplied with $\alpha_D^{1/2}$
    \begin{equation}
        \alpha_D^{1/2}T_{\odot}\sim 10^{3}\left(\frac{\alpha_D}{1}\right)^{1/2}\text{ eV}
    \end{equation}
    which is always above $\omega_D^{\rm th}$ since we always have $T_{\odot}\gg \tilde{T}$.
\end{itemize}
These scales divide the massive dark fermion case into several regimes, each with a different expression for $\Gamma_{2\rightarrow 3}$. We discuss each of these regimes below. The $m_\chi$ dependence of $\Gamma_{2\rightarrow 3}$ and the associated thermalization requirement $\Gamma_{2\rightarrow 3}r_{\rm core}\gtrsim 1$ are illustrated in FIG.~\ref{fig:massivechi}. Note that, in any of these regimes, if $m_\chi\gtrsim \tilde{T}(1\text{ AU})$ dark fermions can thermalize inside the Sun but will fall out of equilibrium (as they become non-relativistic in the fluid frame) before arriving on Earth, hence changing the predicted dark solar wind properties on Earth.

\begin{enumerate}
\item $m_\chi\lesssim \omega_D^{\rm pre}$\\
We begin by reviewing the $m_\chi\ll \omega_D^{\rm pre}$ case in order to highlight the assumptions that led to the $\Gamma_{2\rightarrow 3}\sim \alpha_D^{3/2}\omega_D^{\rm pre}$ expression used in the main text. The sparse high-energy dark fermions $\chi$ produced in the Sun lose their energies dominantly though near-collinear $\chi\chi\rightarrow \chi\chi\gamma_D$ splittings (and their variants). The rate for such $2\rightarrow 3$ splittings 
is naively given by $\alpha_D\Gamma_{2\rightarrow 2}^{\rm soft}$, where $\Gamma_{2\rightarrow 2}^{\rm soft}\sim \alpha_D^2/k_{\rm min}^2$ is the rate of elastic fermion scatterings that is infrared-regulated by the softest dark photon momentum scale, $k_{\rm min}\sim \omega_D^{\rm pre}$. However, the actual rate is suppressed by the so called LPM effect, which is due to the finite time $t_{\rm form}$ it takes to resolve the emitted dark photon (daughter) from the emitting dark fermion (parent). Dark photons emitted within $t_{\rm form}$ interfere destructively and so only one dark photon can be emitted effectively during $t_{\rm form}$. The splitting rate can therefore be written as
\begin{equation}
    \Gamma_{\rm 2\rightarrow 3}\sim \alpha_D\text{ min}\left(\Gamma_{2\rightarrow 2}^{\rm soft},t_{\rm form}^{-1}\right),\quad \text{for}\quad m_\chi\lesssim \omega_D^{\rm pre}
\end{equation}
The formation time $t_{\rm form}$ can be understood as the time it takes for the daughter particle to acquire an $O(1)$ phase in the parent's frame. This is especially relevant in our case because the separation between the parent and daughter particles relies on either their same-direction velocity difference $1-v_\chi\sim (m_\chi/E_\chi)^2$, which is suppressed for a highly relativistic parent $\chi$, or their relative transverse velocity, which is (initially) suppressed due to the collinearity of the emission. In an abelian plasma, the parent-daughter relative transverse velocity is dominated by the deflection of the parent fermion\footnote{By constrast, in a non-abelian plasma, the parent-daughter relative transverse velocity is dominated by the deflection of the daughter gauge boson, which can scatter with other gauge bosons in the plasma owing to their non-abelian nature \cite{Kurkela:2011ti}.}, whose transverse momentum grows as $p_\perp\sim \sqrt{N}\omega_D^{\rm pre}$ as it undergoes $\Delta p_\perp\sim \omega_D^{\rm pre}$ soft collisions $N\sim \Gamma_{2\rightarrow 2}^{\rm soft}t_{\rm form}$ times in the medium. Taking into account both effects, the reciprocal formation time is given by \cite{Peigne:2008wu,Garny:2018grs}
\begin{equation}
    t_{\rm form}^{-1}\sim k\, \text{max}\left[\sqrt{\frac{\alpha_D^2n_{\rm hard}}{E_{\rm hard}^2k}},\left(\frac{m_\chi}{E_{\rm hard}}\right)^2\right]
\end{equation}
The first factor inside the square brackets corresponds to $t_{\rm form}$ being determined by the transverse velocity difference and is the relevant one in the main text. The second factor accounts for the same-direction velocity difference, which does not become important until $m_\chi\gtrsim \omega_D^{\rm th}$. Furthermore, as along as $m_\chi\lesssim \alpha_D^{1/2} T_{\odot}$ the rate $\Gamma_{2\rightarrow 3}$ is set by $t_{\rm form}^{-1}$ which is maximized for $k\sim E_{\rm hard}$, resulting in the following expression used in the main text 
\begin{equation}
    \Gamma_{2\rightarrow 3}\sim \alpha_D t_{\rm form}^{-1}\sim \alpha_D^{3/2}\omega_D^{\rm pre},\quad \text{for}\quad m_\chi\lesssim \omega_D^{\rm pre}
\end{equation}

\item $\omega_D^{\rm pre}\lesssim m_\chi\lesssim\omega_D^{\rm th}$\\
Finite-mass effects begin to matter when $m_\chi$ crosses the pre-thermalization Debye frequency $\omega_{D}^{\rm pre}$ shown in \eqref{kmindebye}. When $m_\chi$ is above this scale, the squared amplitude for the $\gamma_D$ bremsstrahlung of a scattered $\chi$ particle is effectively suppressed by a factor of $(\omega_D^{\rm pre}/m_\chi)^2$, and consequently the $2\rightarrow 3$ rate is now modified as \cite{Peigne:2008wu}
\begin{equation}
    \Gamma_{\rm 2\rightarrow 3}\sim \left[\alpha_D\left(\frac{\omega_D^{\rm pre}}{m_\chi}\right)^2\right]t_{\rm form}^{-1}\sim \left(\frac{\omega_D^{\rm pre}}{m_\chi}\right)^2\alpha_D^{3/2}\omega_D^{\rm pre},\quad \text{for}\quad \omega_D^{\rm pre}\lesssim m_\chi\lesssim \omega_D^{\rm th}    
\end{equation}
This suppression can be traced back to suppression in the emission amplitude, known as the dead cone effect \cite{Dokshitzer:2001zm}, for dark photon emission angles $\theta\lesssim m_\chi/E_{\rm hard}$.

\item $\omega_D^{\rm th}\lesssim m_\chi\lesssim \alpha_D^{1/2} T_{\odot}$\\
The next scale that $m_\chi$ crosses is $(\alpha_D^2n_{\rm hard}E_{\rm hard})^{1/4}$ which is parameterically equivalent to the post-thermalization Debye frequency. Since we are only concerned with the pre-thermalization state of the dark plasma here, the appearance of $\omega_D^{\rm th}$ in this discussion is just a mathematical coincidence. In this regime, the reciprocal formation time $t_{\rm form}^{-1}$ is set by its finite-mass lower bound, $t_{\rm form}^{-1}\gtrsim k (m_\chi/E_{\rm hard})^2$. This has the effect of parametrically enhancing $\Gamma_{2\rightarrow 3}$ relative to the expression applicable in the $\omega_D^{\rm pre}\lesssim m_\chi\lesssim \omega_D^{\rm th}$ regime. The $\Gamma_{2\rightarrow 3}$ in this case is still maximized at $k\sim E_{\rm hard}$ with the value
\begin{equation}
    \Gamma_{2\rightarrow 3}\sim \left[\alpha_D\left(\frac{\omega_D^{\rm pre}}{m_\chi}\right)^2\right]t_{\rm form}^{-1}\sim \, \left(\frac{m_\chi}{\omega_D^{\rm th}}\right)^2\left(\frac{\omega_D^{\rm pre}}{m_\chi}\right)^2\alpha_D^{3/2}\omega_D^{\rm pre},\quad \text{for}\quad \omega_D^{\rm th}\lesssim m_\chi\lesssim \alpha_D^{1/2} T_{\odot}   
\end{equation}
which is independent of $m_\chi$.

\item $m_\chi\gtrsim \alpha_D^{1/2} T_{\odot}$\\
Then, at $m_\chi\gtrsim \alpha_D^{1/2}T_{\odot}$ the reciprocal formation time $t_{\rm form}^{-1}$ goes above the soft $2\rightarrow 2$ scattering rate $\Gamma_{2\rightarrow 2}^{\rm soft}$, which means the LPM effect stops operating, and
\begin{equation}
    \Gamma_{2\rightarrow 3}\sim \left[\alpha_D\left(\frac{\omega_D^{\rm pre}}{m_\chi}\right)^2\right]\Gamma_{2\rightarrow 2}^{\rm soft}\sim \left(\frac{\alpha_D^{1/2}T_{\odot}}{m_\chi}\right)\left(\frac{\omega_D^{\rm pre}}{m_\chi}\right) \alpha_D^{3/2}\omega_D^{\rm pre},\quad \text{for}\quad m_\chi\gtrsim \alpha_D^{1/2} T_{\odot}
\end{equation}
However, it turns out that we always have $\alpha_D^{1/2}T_{\odot}\gtrsim \Tilde{T}(1\text{ AU})$ in the viable parameter space where the dark sector particles satisfy both the cooling limit and self-thermalization condition inside the Sun. Therefore, if $m_\chi$ was high enough to be in this $m_\chi\gtrsim \alpha_D^{1/2}T_{\odot}$ regime, the steady-state fluid properties inside the Earth's orbit would have changed relative to those derived in the main text.
\end{enumerate}


\end{document}